 \documentclass[manuscript]{aastex}

\usepackage{amsmath,mathtools}
\usepackage[titletoc,title]{appendix}

\shorttitle{PERPENDICULAR DIFFUSION COEFFICIENT WITH ADIABATIC FOCUSING}
\shortauthors{WANG ET AL.}

\begin{document}
              \arraycolsep 0pt

\title{PERPENDICULAR DIFFUSION COEFFICIENT OF COSMIC RAYS IN THE PRESENCE OF
WEAK ADIABATIC FOCUSING}

\author{J. F. Wang\altaffilmark{1}, G. Qin\altaffilmark{2},
Q. M. Ma\altaffilmark{1}, T. Song\altaffilmark{1}, and S. B. Yuan\altaffilmark{1}}
\email{}
\altaffiltext{1}{Research Department of Biomedical Engineering, Institute of
Electrical Engineering, Chinese Academy of Science, Beijing 100190, China;
wangjunfang@mail.iee.ac.cn}
\altaffiltext{2}{School of Science, Harbin Institute of Technology, Shenzhen 518055,
China; qingang@hit.edu.cn}

\begin{abstract}
The influence of adiabatic focusing on particle diffusion is an important topic in
astrophysics and plasma physics.
In the past several authors have explored the influence of along-field adiabatic
focusing on parallel diffusion of charged energetic particles.
In this paper by using the Unified NonLinear Transport (UNLT) theory developed by Shalchi
(SH2010) and the method of He and Schlickeiser (HS2014) we derive a new nonlinear
perpendicular diffusion coefficient for non-uniform background magnetic field.
This formula demonstrates that particle perpendicular diffusion coefficient
is modified by along-field adiabatic focusing.
For isotropic pitch-angle scattering and weak adiabatic focusing limit
the derived perpendicular diffusion coefficient
is independent of the sign of adiabatic focusing characteristic length.
For two-component model
we simplify the perpendicular diffusion coefficient up to second order
of the power series of adiabatic focusing characteristic quantity.
We find that the first order modifying factor is equal to zero and the sign of the second
one is determined by the energy of particles.
\end{abstract}

\keywords{perpendicular diffusion coefficient, UNLT, adiabatic focusing}

\section{INTRODUCTION}
Energetic charged particles' propagating through magnetized plasmas is an important
topic in the study of modulation of galactic cosmic rays (GCRs) in heliosphere
\citep[e.g.,][]{Parker63, ZhaoEA14}, solar energetic particles (SEPs) transport in
solar wind
\citep[e.g.,][]{Parker65, Roelof1969, NgAReames94, QinEA05}, energetic particles'
shock acceleration in interplanetary space or from supernova remnants
\citep[e.g.][]{ZankEA00, FerrandEA14}, etc.
According to observations the magnetic field configuration of magnetized plasmas
can be understood as the
superposition of a large-scale background magnetic field $\vec{B}_0=B_0 \hat{e}_z$
and a
turbulent component $\delta\vec{B}$ in many scenarios.
Both the mean magnetic field and the turbulent component can affect the
properties of particles' propagation.
Since the large-scale magnetic field breaks the symmetry of magnetized plasmas,
particles' transport has to be distinguished between parallel and perpendicular
components relative to the mean magnetic field direction.
For simplicity purpose one usually assumes the mean magnetic fields uniform
\citep[e.g.,][]{Schlickeiser2002, Shalchi2009}. However, different observations,
e.g.,
radio continuum surveys of interstellar space and direct in-situ measurements in
solar
system, show the spatially varying large scale magnetic fields for many
scenarios,
which leads to the adiabatic focusing effect of charged energetic particles and
introduces modification to the particle diffusion coefficients
\citep{Roelof1969, Earl1976, Shalchi2011a,  Litvinenko2012a,
Litvinenko2012b, ShalchiEA2013, HeEA2014, WangAQin16}.

For weak adiabatic focusing limit and distribution function with small
anisotropy, along-field adiabatic focusing gives rise to convective transport along and across
the guide field, as well that as in momentum space
\citep{SchlickeiserEA2008}.
In addition, by considering the spatial gradient and curvature of mean magnetic
field, \citet{SchlickeiserEA2010} explored the influence of adiabatic focusing
effects on particle perpendicular propagation. Since only weak limit is considered,
the influence of along-field adiabatic focusing on particle parallel and
perpendicular diffusion coefficient was not found.

By using the Unified NonLinear Transport (UNLT) theory \citep{Shalchi2010}
for arbitrary anisotropic
distribution function, \citet{Shalchi2011a} derived a parallel
diffusion coefficient formula with along-field adiabatic focusing. Employing the
stochastic differential equations, completely equivalent to the Fokker-Planck
equation, another version of parallel diffusion coefficient formula with
adiabatic focusing was obtained \citep{Litvinenko2012a}. This
formula is also obtained in other papers \citep{BeeckEA1986, BieberEA1990, Kota2000,
Litvinenko2012a, HeEA2014}. Analytical and numerical studies
demonstrated that the two versions are related by each other \citep{ShalchiEA2013}.

Analogous to parallel diffusion, perpendicular diffusion is another important
particle transport process. In the past, some  approaches have been developed to
derive the perpendicular diffusion coefficient, e.g., the NonLinear Guiding Center
(NLGC) theory \citep{MatthaeusEA2003}, the Weakly NonLinear Theory (WNLT)
\citep{ShalchiEA2004},
the Modification of NLGC \citep{QinEA2014}, and the Unified NonLinear
Transport (UNLT) theory \citep{Shalchi2010} and so on. However, these theories for
perpendicular diffusion coefficient are all only applicable to the uniform mean
magnetic field and adiabatic focusing effect is not considered.

In summary, a lot of studies have been done for perpendicular diffusion in uniform
mean magnetic field, and parallel diffusion with along-field adiabatic focusing is
also extensively investigated so far. But it is always an open topic whether
along-field adiabatic focusing has influence on perpendicular diffusion.
The perpendicular diffusion coefficient formula for uniform mean magnetic field
derived in \citet{Shalchi2010} (hereafter SH2010) is a well-known result and agrees well with
simulations \citep[e.g.,][]{TautzEA2011, Shalchi2013, HusseinEA2014, ShalchiEA2014}.
In this paper by employing the UNLT theory
and
the method of \citet{HeEA2014} (hereafter HS2014) we explore the perpendicular
diffusion
coefficient with along-field adiabatic focusing effect.

The paper is organized as follows: In the Section 2 we introduce the definitions of
perpendicular diffusion coefficient and along-field adiabatic focusing
length, and from the standard Fokker-Planck equation by employing the
UNLT theory we obtain the equations of the quantities which are related to spatial
perpendicular diffusion coefficient. In Section 3 by using the method of HS2014 we
derive the anisotropic distribution function and obtain the quantities related to
spatial perpendicular
diffusion coefficient which occur in equations in Section 2.
And in Section 4 the
spatial perpendicular diffusion coefficient with along-field adiabatic
focusing is deduced, and for isotropic scattering and weak focusing limit
the first and
second-order modifying factors of perpendicular diffusion coefficient
are derived.
In Section 5 for two-component model
the first and second modifying factors
are explored.
Section 6 shows discussion and conclusions.

\section{SIMPLIFICATION OF THE STANDARD FOKKER-PLANCK EQUATION}
Similar as in SH2010, our starting point is
the single particle motion equation
\citep{MatthaeusEA2003}
\begin{equation}
v_x (t)=av_z (t) \frac{\delta B_x [\vec{x}(t),t]}{B_0 [z(t)]},
	\label{motion equation of single particle motion}
\end{equation}
here $ B_0[ z(t)]$ is the large-scale magnetic field
which are not necessarily uniform,
$\delta B_x$ is the x-component of random magnetic field,
$v_x(t)$ and $v_z(t)$ are the $x$ and $z$ components of particle
speed $v$ respectively.
For convenience, we specialize to transverse fluctuations
$\vec{B_0}\cdot \delta \vec{B}=0$ in this paper.
Because the energetic charged particle feels
the magnetic field at its own position,
so variable
$z$
in the mean magnetic field $B_0(z)$ is a random variable with time.
The parameter $a^2$ indicates the relation between single particle and the
magnetic field lines, and the relevant details can be found in the papers of NLGC
and UNLT theories and so on \citep{MatthaeusEA2003, Shalchi2015, Shalchi2016}.

In this paper the well-known TGK (Taylor-Green-Kubo) formulation
\citep[see,][]{Taylor1922, Green1951, Kubo1957} is used to compute perpendicular diffusion
coefficient
\begin{equation}
\kappa_\bot=\int_{0}^{\infty}dt \langle v_x (t) v_x (0) \rangle.
	\label{Kubo formula}
\end{equation}

\subsection{Along-field adiabatic focusing length}
The along-field adiabatic focusing length for spatially varying
guiding field is shown as \citep{SchlickeiserEA2008}
\begin{equation}
L^{-1}(z)=-\frac{1}{B_{0}(z)}\frac{dB_{0}(z)}{dz}.
\end{equation}

For the limit $L \rightarrow{\infty}$ the spatially varying background magnetic field
$B_0 (z)$ tends
to be uniform.
In this paper, for mathematical tractability we assume the adiabatic focusing length
is a constant, i.e., $\partial{L} / \partial{z}=0 $.
Then the mean magnetic field can be written as
\begin{equation}
B_0 (z)=B_0 (0) e^{-z/L}.
\label{The mean magnetic field with exponential function}
\end{equation}

For convergent background magnetic field along $z$ positive direction adiabatic
focusing length satisfies $L<0$, and for divergent case $L>0$.

\subsection{The formula of perpendicular diffusion coefficient}
By combining Equations (\ref{motion equation of single particle motion}) and
(\ref{Kubo formula}), perpendicular diffusion coefficient $\kappa_\bot$ can be
obtained
\begin{equation}
\kappa_\bot=\frac{a^2 v^2}{B_{0}^{2}(0)}\int d^3 k P_{xx}(\vec{k})T(\vec{k})
	\label{Originally perpendicular diffusion coefficient with T}
\end{equation}
with
\begin{equation}
\begin{aligned}
T(\vec{k})&=\frac{1}{v^2}\int_{0}^{\infty} dt \langle v_z (t) v_z (0)
e^{z(t)/L}e^{i\vec{k}\cdot\vec{x(t)}}\rangle\\
&=\frac{1}{4}\int_{0}^{\infty}dt
\int_{-1}^{+1} d\mu \mu \int_{-1}^{+1} d\mu_0 \mu_0 \int d^3 x e^{i\vec{k}
\cdot\vec{x}(t)}e^{z(t)/L}f_0 [\vec{x}(t), \mu, \mu_0, t],
\label{Definition of T}
\end{aligned}
\end{equation}
here $\mu$ is pitch-angle cosine and $\mu_0$ is initial pitch-angle cosine,
$P_{xx}(\vec{k})$ is  the magnetic field correlation tensor.
In Equation (\ref{Originally perpendicular diffusion coefficient with T})
the Corrsin independence hypothesis \citep[see,][]{Corrsin1959}
is employed.
And the quantity $T(\vec{ k})$ is required to be real.
Note that we also use the standard assumptions
of magnetic turbulence,
e.g., axisymmetry, homogeneity, magnetostatic, and so on.
And $f_0 [\vec{x}(t), \mu, \mu_0, t]$ in Equation
(\ref{Definition of T})
is the distribution function of the standard Fokker-Planck equation
with adiabatic focusing effect,
which is shown as follow
\begin{equation}
\frac{\partial{f_0}}{\partial{t}}+v\mu \frac{\partial{f_0}}{\partial{z}}=
\frac{\partial{}}{\partial{\mu}} \left(D_{\mu\mu}
\frac{\partial{f_0}}{\partial {\mu}}\right)-\frac{v(1-\mu^2)}{2L}
\frac{\partial{f_0}}{\partial{\mu}} +D_\bot \Delta_\bot f_0.
	\label{standard Fokker-Planck equation}
\end{equation}
The latter equation includes
the Fokker-Planck coefficients of pitch-angle
diffusion $D_{\mu \mu}$ and the perpendicular diffusion $D_\bot$,
along-field adiabatic focusing length $L$, and the differential
operator $\Delta_\bot =\partial ^2 / \partial x^2+\partial ^2 / \partial y^2$.
However, the source term is neglected.

Equation (\ref{Definition of T}) can also be written as
\begin{equation}
T(\vec{k})=\frac{1}{2}\int_{-1}^{1} d\mu \mu S(\vec{k},\mu), \label{T}
\end{equation}
with the quantity $S(\vec{k},\mu)$
as
\begin{equation}
S(\vec{k},\mu)=\frac{1}{2}\int_{0}^{\infty}dt \int_{-1}^{1} d\mu_0 \mu_0 \int d^3x
e^{i\vec{k}\cdot\vec{x}}e^{z(t)/L}f_0 [\vec{x}(t), \mu, \mu_0, t].
\label{S with f}
\end{equation}

From Equations
(\ref{Originally perpendicular diffusion coefficient with T}), (\ref{T}) and (\ref{S with f}) we
can find
that in order to get perpendicular diffusion coefficient
$\kappa_\bot$ the formula $S(\vec{k},\mu)$ has to be obtained.

\subsection{The governing equation of $\boldsymbol{S(\vec{k},\mu)}$}
Let's  set $f [\vec{x}(t), \mu, \mu_0, t]
=e^{z(t)/L}f_0 [\vec{x}(t), \mu, \mu_0, t]$
and combining the standard Fokker-Planck equation
(see, Equation (\ref{standard Fokker-Planck equation}))
we can obtain the following equation
\begin{equation}
\frac{\partial{f}}{\partial{t}}+v\mu \frac{\partial{f}}{\partial{z}}=
\frac{\partial{}}
{\partial{\mu}} \left[D_{\mu\mu} \frac{\partial{f}}{\partial {\mu}}-\frac{v}{2L}
(1-\mu ^2) f \right]+\frac{v\mu}{L}f+D_\bot \Delta_\bot f,
	\label{Modified Fokker-Planck equation}
\end{equation}

In what follows, from the latter equation we use the UNLT theory to
derive the governing equation of $S(\vec{k},\mu)$.
 By employing the relation
$f [\vec{x}(t), \mu, \mu_0, t]=e^{z(t)/L}f_0 [\vec{x}(t), \mu, \mu_0, t]$
the formula of quantity $S(\vec{k}, \mu)$ (see, (\ref{S with f}))
can be rewritten as
\begin{equation}
S(\vec{k},\mu)=\frac{1}{2}\int_{0}^{\infty}dt \int_{-1}^{1}
d\mu_0 \mu_0 \Gamma [\vec{k}(t), \mu, \mu_0, t].
\label{S with f-2}
\end{equation}
with
\begin{equation}
\Gamma [\vec{k}(t), \mu, \mu_0, t]=\int d^3 x
e^{i\vec{k}\cdot\vec{x}}f [\vec{x}(t), \mu, \mu_0, t].
\end{equation}

After operating Fourier transformation on Equation
(\ref{Modified Fokker-Planck equation})
over spatial variable $\vec{x}$
we can get
\begin{equation}
\frac{\partial{\Gamma}}{\partial{t}}-i k_\parallel v\mu \Gamma =\frac{\partial{}}
{\partial{\mu}} \left[D_{\mu \mu }\frac{\partial{\Gamma}}{\partial {\mu}}-\frac{v}
{2L}\left(1-\mu ^2 \right)\Gamma \right]+\frac{v\mu}{L}\Gamma
-k_\bot ^2 D_\bot \Gamma,
\label{Taking Fourier transformation to Fokker-Planck equation}
\end{equation}
where we assume that $D_{\mu\mu}$ and $D_\bot$ are independent of
$\vec{x}$.

To multiply the time integral of Equation
(\ref{Taking Fourier transformation to Fokker-Planck equation}) from 0 to $\infty$
with the initial pitch-angle cosine $\mu_{0}$, and then integrate the results over
$\mu_{0}$ from $-1$ to $1$, finally we can obtain
\begin{eqnarray}
\begin{aligned}
&\int_{-1}^{1}d \mu_{0} \mu_{0} \Gamma (t=\infty) -\int_{-1}^{1}d\mu_{0} \mu_{0}
\Gamma (t=0)-i k_\parallel v\mu \int_{-1}^{1}d\mu_{0}
\mu_{0}\int_{0}^{\infty}d t \Gamma \\
&=\frac{\partial{}}{\partial{\mu}}\left[D_{\mu \mu} \frac{\partial{}}{\partial{\mu}}
\int_{-1}^{1}d\mu_{0} \mu_{0}\int_{0}^{\infty}d t \Gamma -\frac{v}{2L}(1-
\mu ^2)\int_{-1}^{1}d\mu_{0}\mu_{0}\int_{0}^{\infty}d t \Gamma \right]\\
&+\frac{v\mu}{L}\int_{-1}^{1}d\mu_{0}\mu_{0}\int_{0}^{\infty}d t \Gamma
-k_\bot ^2 D_\bot \int_{-1}^{1}d\mu_{0}\mu_{0}\int_{0}^{\infty}d t \Gamma.
\end{aligned}
\end{eqnarray}

The initial distribution can be defined as
\begin{equation}
\Gamma (t=0)=2\delta(\mu-\mu_0)h(\vec{k}_0)
=2\delta(\mu-\mu_0)\int  d^3 x_0 e^{i\vec{k}_0\cdot
	\vec{x}_0}h(\vec{x}_0),
  \label{Original condition for t=0}
\end{equation}
which denotes that the particles have a well-defined initial pitch-angle cosine,
and $h(\vec{k}_0)$ is the distribution function in phase space
for initial time $t=t_0$.
By assuming $\Gamma(t\rightarrow \infty)$
to be independent of the initial pitch-angle cosine
$\mu_0$ as in \citet{Shalchi2010}, we can obtain the following equation
\begin{eqnarray}
\begin{aligned}
&-2\mu h(\vec{k}_0)-i k_\parallel v\mu \int_{-1}^{1}d\mu_{0}\mu_{0}\int_{0}^{\infty}d
t\Gamma \\
&= \frac{\partial{}}{\partial{\mu}} \left[D_{\mu \mu} \frac{\partial{}}
	{\partial{\mu}}
\int_{-1}^{1}d \mu_{0}\mu_{0} \int_{0}^{\infty} d t\Gamma -\frac{v}{2L}
	\left(1-\mu ^2
\right)\int_{-1}^{1}d\mu_{0}\mu_{0} \int_{0}^{\infty} d t \Gamma \right]\\
&+\frac{v \mu}{L}\int_{-1}^{1}d\mu_{0}\mu_{0}\int_{0}^{\infty}d t \Gamma
-k_\bot ^{2}D_\bot \int_{-1}^{1}d\mu_{0}\mu_{0}\int_{0}^{\infty}d t\Gamma,
\label{Fokker-Planck equation with original condition}
\end{aligned}
\end{eqnarray}
here, since $\vec{x}=\vec{x}_0$ for time $t=t_0$, Fourier transformation
$\int d^3 x e^{i\vec{k}\cdot\vec{x}} h(\vec{x})$ becomes
$\int d^3 x_0 e^{i\vec{k}_0\cdot\vec{x}_0} h(\vec{x}_0)$.
And in the latter equation the following formula is used
\begin{equation}
h(\vec{k}_0)=\int d^3 x_0 e^{i\vec{k}_0\cdot\vec{x}_0} h(\vec{x}_0),
\end{equation}
where the normalization condition $\int d^3 k_0 h(\vec{k}_0)=1$ must be satisfied
for the initial time $t=t_0$.

To proceed, by using the formula $S(\vec{k},\mu)
=(1/2)\int_{0}^{\infty}dt \int_{-1}^{1} d\mu_0 \mu_0
\Gamma [\vec{k}(t), \mu, \mu_0, t]$
(see Equation (\ref{S with f-2}))
the governing equation of $S(\vec{k},\mu)$ can be got from
Equation
(\ref{Fokker-Planck equation with original condition})
\begin{equation}
-\mu h(\vec{k}_0)-i k_\parallel v \mu S =\frac{\partial{}}{\partial{\mu}}\left[
D_{\mu \mu}\frac{\partial{S}}{\partial{\mu}}-\frac{v}{2L} \left(1-\mu ^2\right) S
\right]+\frac{v\mu S}{L}-k_\bot ^2 D_\bot S.
\label{eq:muh}
\end{equation}
Splitting $S(\mu,\vec{k})$ into an even function $S_{+}(\mu,\vec{k})$ and an odd
function
$S_{-}(\mu,\vec{k})$, i.e., $S(\mu,\vec{k})=S_{+}(\mu,\vec{k})+S_{-}(\mu,\vec{k})$,
and then inserting this formula into Equation (\ref{eq:muh}) yields
\begin{eqnarray}
\begin{aligned}
&-\mu h(\vec{k}_0)-i k_\parallel v \mu S_{+} -\frac{\partial{}}{\partial{\mu}} \left[
D_{\mu \mu}  \frac{\partial{S_{-}}}{\partial{\mu}}\right]-2\frac{v\mu}{L}S_{+}
+\frac{v}{2L}
\left(1-\mu ^2\right) \frac{\partial{S_{+}}}{\partial{\mu}}+ k_\bot ^2 D_\bot S_{-}\\
&= i k_\parallel v \mu S_{-} +\frac{\partial{}}{\partial{\mu}} \left[D_{\mu \mu}
\frac{\partial{S_{+}}}{\partial{\mu}}\right]+ 2\frac{v\mu}{L}S_{-} -\frac{v}{2L}
	\left(1-\mu
^2\right)\frac{\partial{S_{-}}}{\partial{\mu}}-k_\bot ^2 D_\bot S_{+}.
\label{S+S-}
\end{aligned}
\end{eqnarray}
To proceed, by integrating Equation (\ref{S+S-})
over $\mu$ from -1 to 1 and from 0 to 1 respectively,
thereafter using the standard assumption $D_{\mu \mu} (\mu =\pm 1)=0$
we can get the following integral equations
\begin{equation}
(i k_\parallel v +\frac{v}{L})\int_{-1}^{1} d\mu \mu S_{-}
=k_\bot ^{2} \int_{-1}^{1} d\mu
	D_\bot S_{+}     \label{The first one of equations of S+ and S-}
\end{equation}
and
\begin{equation}
\begin{aligned}
&-\frac{1}{2}h(\vec{k}_0)-(i k_\parallel v +\frac{v}{L})
\int_{0}^{1}d\mu \mu S_{+} +
	\left(D_{\mu \mu}
\frac{\partial{S_{-}}}{\partial{\mu}}\right)(\mu =0) -\frac{v}{2L}S_{+}(\mu=0) +
k_\bot ^2 \int_{0}^{1} d\mu D_\bot S_{-} \\
&=-\left(D_{\mu \mu} \frac{\partial{S_{+}}}{\partial{\mu}}\right)(\mu=0)
	+\frac{v}{2L}S_{-}(\mu=0)
	\label{The second one of equations of S+ and S-}.
\end{aligned}
\end{equation}
If employing pitch-angle isotropic scattering model
$D_{\mu \mu}=D \left(1-\mu ^2\right)$
with a constant $D$ \citep[e.g.,][]{ShalchiEA2009, Shalchi2010} and
the perpendicular diffusion coefficient model
$D_\bot=2\kappa_\bot |\mu|$ \citep[e.g.,][]{QinAShalchi14, Shalchi2010, Shalchi2017}, Equation
(\ref{The second one of equations of S+ and S-}) can
be simplified as
\begin{equation}
\begin{aligned}
&-\frac{1}{2}h(\vec{k}_0)-(i k_\parallel v
+\frac{v}{L}) \int_{0}^{1}d\mu \mu S_{+} +
D \frac{\partial{S_{-}}}{\partial{\mu}}(\mu =0) -\frac{v}{2L}S_{+}(\mu=0) +
2\kappa_\bot k_\bot ^2 \int_{0}^{1}  d\mu \mu S_{-}\\
& =-D \frac{\partial{S_{+}}}{\partial{\mu}}(\mu=0)+\frac{v}{2L}S_{-}(\mu=0)
	\label{Last equation of S+ S-}.
\end{aligned}
\end{equation}

Since uniform mean magnetic field was considered in SH2010, only the two terms
$\partial{S_{-}} / \partial{\mu}(\mu =0)$ and
$\partial{S_{+}} / \partial{\mu}(\mu =0)$
occurred in the governing equations (see Equations (15) and (16) in SH2010).
Since adiabatic focusing effect considered in this paper, we have to determine
the four terms $\partial{S_{-}} / \partial{\mu}(\mu =0)$,
$\partial{S_{+}} / \partial{\mu}(\mu =0)$, $S_+ (\mu=0)$ and $S_- (\mu=0)$ in
Equations
(\ref{The first one of equations of S+ and S-}) and (\ref{Last equation of S+ S-}).
If the approximation formula $\partial{S_{-}} / \partial{\mu}\approx
\partial{S_{-}} / \partial{\mu}(\mu =0)$
in SH2010 and the approximation method in \citet{Shalchi2011b}
are not used in Equations
(\ref{The first one of equations of S+ and S-}) and (\ref{Last equation of S+ S-}),
the specific expression of $S(\vec{k},\mu)$ has to be obtained.

\section{THE FORMULA OF $\boldsymbol{S(\vec{k},\mu)}$}
The distribution function $f(\vec{x},\mu,\mu_0,t)$ of
Equation (\ref{Modified Fokker-Planck equation}) can be split into the isotropic
distribution part $F(\vec{x},t)$ and anisotropic distribution part
$g(\vec{x}, \mu, \mu_{0}, t)$
\begin{equation}
f(\vec{x}, \mu, t)=F(\vec{x},t)+g(\vec{x}, \mu, \mu_{0}, t).      \label{f=F+g}
\end{equation}
By inserting Equation
(\ref{f=F+g}) into the formula of $S(\vec{k},\mu)$ (see Equation (\ref{S with f-2})),
since $F(\vec{x},t)$ does not contain $\mu_0$,
we can get
\begin{equation}
S(\mu,\vec{k})=\frac{1}{2} \int_{-1}^{1}d\mu_{0} \mu_{0} \int_{0}^{\infty}d t
	\int d^3 x e^{i \vec{k}\cdot \vec{x}} g(\vec{x}, \mu, \mu_{0}, t).
	\label{S with g}
\end{equation}
If the anisotropic distribution function $g(\vec{x}, \mu, \mu_{0}, t)$ is obtained,
the specific
expressions of $S(\mu,\vec{k})$, $S_+(\vec{k},\mu)$, and $S_-(\vec{k},\mu)$ can be
obtained from Equation (\ref{S with g}).
In the next Subsection we will employ the method of HS2014 to obtain
$g(\vec{x}, \mu, \mu_{0}, t)$. Consequently, the terms
$\partial{S_{-}} / \partial{\mu}(\mu =0)$,
$\partial{S_{+}} / \partial{\mu}(\mu =0)$, $S_+ (\mu=0)$, and $S_- (\mu=0)$ contained in
Equations
(\ref{The first one of equations of S+ and S-}) and (\ref{Last equation of S+ S-})
will also be obtained.

\subsection{The anisotropic distribution function $\boldsymbol{g(\vec{x}, \mu, \mu_{0}, t)}$}
By using the method of HS2014, from Equation
(\ref{Modified Fokker-Planck equation}) the exact anisotropic distribution function
can be written as
\begin{equation}
g(\mu)=\left[1-\frac{2e^{M(\mu)}}{\int_{-1}^{1}d\mu e^{M(\mu)}} \right] L
\left(\frac{\partial{F}}{\partial{z}}-2\frac{F}
{L}\right)+e^{M(\mu)}\Bigg[\int_{-1}^{\mu}d\nu R(\nu)-\frac{\int_{-1}^{1}d\mu
e^{M(\mu)}\int_{-1}^{\mu}d\nu R(\nu)}{\int_{-1}^{1}d\mu e^{M(\mu)}}\bigg],
\label{g}
\end{equation}
where $R(\mu)$ can be written as
\begin{equation}
R(\mu)=R_1(\mu)+R_2(\mu)+R_3(\mu)
	\label{R=R1+R2+R3}
\end{equation}
with
\begin{eqnarray}
R_1(\mu)&=&\frac{e^{-M(\mu)}}{D_{\mu \mu}}\left(\mu \frac{\partial{F}}
{\partial{t}}+\int_{-1}^{\mu}d\nu \frac{\partial{g}}{\partial{t}}\right)\\
R_2(\mu)&=&\frac{e^{-M(\mu)}}{D_{\mu \mu}}\left( v\frac{\partial{}}
{\partial{z}}-\frac{v}{L}\right)\left(\int_{-1}^{\mu}d\nu \nu
g-\dfrac{1}{2}\int_{-1}^{1}d\mu \mu g
	\right)\\
R_3(\mu)&=&\frac{e^{-M(\mu)}}{2D_{\mu \mu}}\left(\int_{-1}^{1}d
	\mu D_\bot
\Delta_\bot F+\Delta_\bot \int_{-1}^{1}d\mu D_\bot g \right)
	\nonumber\\
&&-\frac{e^{-M(\mu)}}{D_{\mu \mu}}\left(\Delta_\bot
F\int_{-1}^{\mu}d\mu D_\bot+\Delta_\bot \int_{-1}^{\mu}d\mu D_\bot g \right),
\label{R3}
\end{eqnarray}
and the quantity $M(\mu)$ is introduced as in HS2014
\begin{equation}
M(\mu)=\frac{v}{2L}\int_{-1}^{\mu}d\nu\frac{1-\nu^2}{D_{\nu\nu}(\nu)}.
\end{equation}
For pitch-angle isotropic scattering $D_{\mu \mu}=D(1-\mu^2)$ the quantity $M(\mu)$ can be
simplified as
\begin{equation}
M(\mu)=\xi (1+\mu)
 \end{equation}
with $\xi=v / (2DL)$.
From Equations (\ref{g})-(\ref{R3}) we can see that anisotropic distribution
function $g(\mu)$ is an iteration function by itself. In other words, $R(\mu)$ is an
iteration function by itself.

\subsection{The formula of $\boldsymbol{S(\vec{k},\mu)}$}
In Subsection 3.1 the anisotropic distribution function $g(\mu)$ is obtained by
using the
method of HS2014.
In what follows, we derive the specific formula of $S(\vec{k},\mu)$, and then obtain the
expressions of $S_+ (\vec{k},\mu)$ and $S_-(\vec{k},\mu)$
which are the keys to express the perpendicular diffusion coefficient.

Inserting anisotropic distribution function (\ref{g}) into Equation (\ref{S with g})
we can get
\begin{equation}
\begin{aligned}
S(\vec{k},\mu)&=\frac{1}{2}e^{M(\mu)} \int_{-1}^{1}d\mu_{0} \mu_{0}
	\int_{0}^{\infty}d t
\int d^3 x e^{i \vec{k}\cdot \vec{x}} \int_{-1}^{\mu}d\nu R(\nu)\\
&-\frac{1}{2}\frac{e^{M(\mu)}}{\int_{-1}^{1}d\mu e^{M(\mu)}} \int_{-1}^{1}d\mu
e^{M(\mu)}\int_{-1}^{1}d\mu_{0} \mu_{0} \int_{0}^{\infty}d t \int d^3 x e^{i
\vec{k}\cdot \vec{x}} \int_{-1}^{\mu}d\nu R(\nu),
    \label{S by R}
\end{aligned}
\end{equation}
from which we can see that in order to get the expression for $S(\vec{k},\mu)$ we
have to obtain the results of
$\int_{-1}^{1}d\mu_{0} \mu_{0} \int_{0}^{\infty}d t \int d^3 x e^{i \vec{k}\cdot
\vec{x}}
\int_{-1}^{\mu}d\nu R(\nu)$. By assuming the pitch-angle isotropic scattering
$D_{\mu\mu}=D(1-\mu^2)$ and using Equations (\ref{R=R1+R2+R3})-(\ref{R3}) the
following formula can be obtained
\begin{equation}
\begin{aligned}
&\int_{-1}^{1}d\mu_{0} \mu_{0} \int_{0}^{\infty}d t \int d^3 x e^{i \vec{k}\cdot
\vec{x}} \int_{-1}^{\mu}d\nu R(\nu)
=\frac{1}{D}\int_{-1}^{\mu}d\nu e^{-M(\nu)}\int d^3 x_{0}
e^{i\vec{k}_0\cdot\vec{x}_0}h(\vec{x}_0)+\Phi(\vec{k},\vec{k}_0,\mu),
\label{First-order R with Phi(k0)}
\end{aligned}
\end{equation}
with
\begin{equation}
\begin{aligned}
&\Phi(\vec{k},\vec{k}_0,\mu)=\int_{-1}^{1}d\mu_{0} \mu_{0} \int_{0}^{\infty}d t
\int_{-1}^{\mu}d\nu \frac{e^{-M(\nu)}}{D_{\nu \nu}} \Bigg[\frac{1}{2}\left( -ik_z v
-\frac{v}{L}\right)\left(2\int_{-1}^{\nu}d\rho g(\vec{k})
-\int_{-1}^{1}d\mu \mu g(\vec{k})\right)\\
&+\frac{k_\bot ^2}{2}\left(2\int_{-1}^{\nu}d\rho D_\bot g(\vec{k})-\int_{-1}^{1}d
	\mu D_\bot g(\vec{k})\right)\Bigg],
    \label{Phi}
\end{aligned}
\end{equation}
where the formula $g(\vec{k})=\int d^3 x e^{i\vec{k}\cdot\vec{x}}g(\vec{x})$
and the original condition (\ref{Original condition for t=0}) are used.

Combining Equations (\ref{S by R}) and (\ref{First-order R with Phi(k0)}) gives
\begin{equation}
S(\vec{k},\mu)=\frac{1}{2D}e^{M(\mu)}\left[\int_{-1}^{\mu}d\nu
	e^{-M(\nu)}-
\frac{\int_{-1}^{1}d\mu e^{M(\mu)}\int_{-1}^{\mu}d\nu e^{-M(\nu)}}{\int_{-1}^{1}d
	\mu e^{M(\mu)}}\right]h(\vec{k}_0)+\Lambda(\vec{k},\vec{k}_0,\mu),
    \label{S with Lambda(k0)}
\end{equation}
with
\begin{equation}
\Lambda(\vec{k},\vec{k}_0,\mu)=\frac{1}{2}e^{M(\mu)}\left[\Phi(\vec{k},\vec{k}_0,
	\mu)-
\frac{\int_{-1}^{1}d\mu e^{M(\mu)}\Phi(\vec{k},\vec{k}_0,\mu)}{\int_{-1}^{1}d\mu
e^{M(\mu)}}\right], \label{Lambda}
\end{equation}
here $\Phi(\vec{k},\vec{k}_0,\mu)$ is Equation (\ref{Phi}),
and from Equations (\ref{S with Lambda(k0)}) and
(\ref{Lambda}) we can find that $S(\vec{k},\mu)$ contains variable $\vec{k_0}$, i.e.,
$S(\vec{k},\mu)=S(\vec{k},\vec{k_0},\mu)$.
From Equations (\ref{g}), (\ref{Phi}),
(\ref{S with Lambda(k0)}) and (\ref{Lambda}) we can see that
in order to get the formula of $S(\vec{k},\mu)$
we have to explore the expression of $R(\mu)$.
Since $R(\mu)$ is an infinite iteration
expression
by itself, so $\Phi(\vec{k},\vec{k}_0,\mu)$ is
the infinite iteration function of $R(\mu)$.
Therefore, $\Lambda(\vec{k},\vec{k}_0,\mu)$
as well as $S(\vec{k},\vec{k}_0,\mu)$
is an infinite iteration function of
$R(\mu)$. Using the assumption $D_{\mu\mu}=D(1-\mu^2)$
Equation (\ref{S with Lambda(k0)}) can be simplified as
\begin{equation}
S(\vec{k},\vec{k}_0,\mu)=\frac{1}{2D}\frac{\xi \cosh (\mu \xi)+\xi \sinh(\mu \xi)-
	\sinh \xi}{\xi \sinh \xi} h(\vec{k}_0)+\Lambda(\vec{k},\vec{k}_0,\mu).
    \label{23-2}
\end{equation}

\subsection{The formulas of $\boldsymbol{S_+ (\vec{k},\vec{k}_0,\mu)}$ and
$\boldsymbol{S_- (\vec{k},\vec{k}_0,\mu)}$}
The quantity $S(\vec{k}, \mu)$ can be split as the sum of the even function
$S_+ (\vec{k},\mu)$ and the odd function $S_- (\vec{k},\mu)$ of $\mu$,
and $S_+ (\vec{k},\mu)$ and $S_- (\vec{k},\mu)$ is shown as follows
\begin{eqnarray}
S_+ (\vec{k},\vec{k}_0,\mu)&=& \frac{\xi \cosh (\mu \xi) -\sinh \xi}{2D\xi \sinh
\xi}h(\vec{k}_0)+\Lambda_+ (\vec{k}, \vec{k}_0, \mu),  \label{S+ with Lambda+}\\
S_- (\vec{k},\vec{k}_0,\mu)&=& \frac{ \sinh (\mu \xi)}{2D \sinh
\xi}h(\vec{k}_0)+\Lambda_-
(\vec{k}, \vec{k}_0, \mu),
	\label{S- with Lambda-}
\end{eqnarray}
where we use the even function $\Lambda_+ (\vec{k}, \vec{k}_0, \mu)$ and the odd
function
$\Lambda_- (\vec{k}, \vec{k}_0, \mu)$ of $\mu$ which satisfy the formula
$\Lambda (\vec{k}, \vec{k}_0, \mu)=\Lambda_+ (\vec{k}, \vec{k}_0, \mu)
+\Lambda_- (\vec{k}, \vec{k}_0, \mu)$.
Obviously, $S_+(\vec{k}, \vec{k}_0, \mu)$ and
$S_-(\vec{k}, \vec{k}_0, \mu)$ are all iteration function of $R(\mu)$.

By using formulas (\ref{S+ with Lambda+}) and (\ref{S- with Lambda-}) we can get the
following quantities
\begin{eqnarray}
S_+ (\vec{k},\vec{k}_0,\mu=0)&=& \frac{\xi -\sinh \xi}{2D\xi \sinh
\xi}h(\vec{k}_0)+\Lambda_+ (\vec{k}, \vec{k}_0, \mu=0),
\label{S+(mu=0)}\\
S_- (\vec{k},\vec{k}_0,\mu=0)&=& 0,\\
\frac{\partial{S_+}}{\partial{\mu}}(\vec{k},\vec{k}_0,\mu=0)&=& 0,\\
\frac{\partial{S_-}}{\partial{\mu}}(\vec{k},\vec{k}_0,\mu=0)&=& \frac{ \xi}{2D\sinh
\xi}h(\vec{k}_0)+\frac{\partial{\Lambda_- (\vec{k}, \vec{k}_0, \mu)}}{\partial{\mu}}
(\mu=0).
\label{ds-/dmu(mu=0)}
\end{eqnarray}

From Equation (\ref{S- with Lambda-}) the following quantity can be obtained
\begin{equation}
\frac{\partial{S_-}(\vec{k},\vec{k}_0,\mu)}{\partial{\mu}}= \frac{\xi\cosh (\mu\xi)}
	{2D
\sinh \xi}h(\vec{k}_0)+\frac{\partial{\Lambda_- (\vec{k}, \vec{k}_0, \mu)}}
{\partial{\mu}}.
\end{equation}

The formula $\Lambda_- (\vec{k}, \vec{k}_0, \mu)$ is very complicated, and if
$\partial \Lambda_- (\vec{k}, \vec{k}_0, \mu)/\partial \mu$ can be neglected, the
following approximation can be obtained
\begin{equation}
\frac{\partial{S_-}(\vec{k},\vec{k}_0,\mu)}{\partial{\mu}}\approx\frac{\partial{S_-}
(\vec{k},\vec{k}_0,\mu)}{\partial{\mu}}(\mu=0),    \label{eq:pSapprox}
\end{equation}
here weak adiabatic focusing limit $\cosh (\mu\xi)\approx 1$  is used. Note that the
Equation (\ref{eq:pSapprox}) is precisely the approximation used in SH2010.

\section{THE PERPENDICULAR DIFFUSION COEFFICIENT WITH ALONG-FIELD ADIABATIC FOCUSING}
Next, Equation (\ref{T}) can be rewritten as
\begin{equation}
T(\vec{k},\vec{k}_0)=\frac{1}{2}\int_{-1}^{1} d\mu\mu S_-(\vec{k}, \vec{k}_0,
	\mu).
\end{equation}
By considering Equation (\ref{S- with Lambda-}), $T(\vec{k},\vec{k}_0)$ becomes
\begin{equation}
 T(\vec{k},\vec{k}_0)= \frac{\xi \cosh (\mu \xi) -\sinh \xi}{2D\xi^2 \sinh
 \xi}h(\vec{k}_0)+\frac{1}{2}\int_{-1}^{1} d\mu\mu \Lambda_- (\vec{k}, \vec{k}_0,
	\mu).
 \label{T with HS2014}
\end{equation}

In fact, using Equations
(\ref{Originally perpendicular diffusion coefficient with T}) and
(\ref{T with HS2014}) the perpendicular
diffusion coefficient can also be obtained,
but we want to leave this for future work.

Equation (\ref{T with HS2014}) can be rewritten as
\begin{equation}
 T(\vec{k},\vec{k}_0)=\left[\frac{\xi \cosh (\mu \xi) -\sinh \xi}{2D\xi^2 \sinh
 \xi}+\frac{1}{h(\vec{k}_0)}\int_{0}^{1} d\mu\mu \Lambda_- (\vec{k}, \vec{k}_0, \mu)
 \right]h(\vec{k}_0).
	\label{T-h(k0)}
\end{equation}

Inserting Equations (\ref{S+(mu=0)})-(\ref{ds-/dmu(mu=0)}) and  (\ref{T-h(k0)}) into
Equations (\ref{The first one of equations of S+ and S-}) and
(\ref{Last equation of S+ S-}) we can get
\begin{equation}
T(\vec{k},\vec{k}_0)= \frac{1}{3}\frac{1}{(4/3)\kappa_\bot k_\bot ^2
-(ik_\parallel v +2D\xi)^2 / (3\kappa_\bot k_\bot^2)
+\Xi(\vec{k}, \vec{k}_0, \xi)  }h(\vec{k}_0)
\label{T with h(k0)}
\end{equation}
with
\begin{equation}
\Xi(\vec{k}, \vec{k}_0, \xi)= \frac{2D}{3}\frac{1+[2D/h(\vec{k}_0)]
	(\partial{\Lambda_-
(\vec{k}, \vec{k}_0, \mu)}/\partial{\mu})(\mu=0)-2D\xi \Lambda_+
	(\vec{k}, \vec{k}_0,
\mu=0)/h(\vec{x}_0 )}{(\xi\cosh \xi -\sinh \xi)/(\xi^2 \sinh
\xi)+2D\int_{0}^{1}d\mu\mu\Lambda_- (\vec{k},\mu)/h(\vec{k}_0)},
\label{accurate iteration function}
\end{equation}
where the perpendicular diffusion coefficient model
$D_\bot =2\kappa_\bot |\mu|$ and $2D\xi=v/L$ are
employed.

Integrating Equation (\ref{T with h(k0)}) over $\vec{k}_0$ for all values and then
inserting it into Equation
(\ref{Originally perpendicular diffusion coefficient with T}),
the following equation can be obtained
\begin{equation}
\kappa_\bot =\frac{a^2 v^2}{3B_0 ^2 (0)}\int d^3 k\int d^3 k_0
\frac{ P_{xx}(\vec{k})h(\vec{k}_0)}{(4/3)\kappa_\bot k_\bot ^2
-(ik_\parallel v +2D\xi)^2 / (3\kappa_\bot k_\bot^2)
+\Xi(\vec{k}, \vec{k}_0, \xi)}.
\label{Accurate perpendicular diffusion coefficient}
\end{equation}
The latter equation can be rewritten as
\begin{equation}
\begin{aligned}
&\kappa_\bot =\frac{a^2 v^2}{3B_0 ^2 (0)}
\int d^3 k\int d^3 k_0 P_{xx}(\vec{k})h(\vec{k}_0)\\
&\frac{(4/3)\kappa_\bot k_\bot ^2+(k_\parallel v)^2 / (3\kappa_\bot k_\bot^2)
-(2D)^2\xi ^2 / (3\kappa_\bot k_\bot^2)
+\Xi(\vec{k}, \vec{k}_0, \xi)+i(4vDk_\parallel \xi)/ (3\kappa_\bot k_\bot^2)}
{[(4/3)\kappa_\bot k_\bot ^2+(k_\parallel v)^2 / (3\kappa_\bot k_\bot^2)
-(2D)^2\xi ^2 / (3\kappa_\bot k_\bot
^2)+\Xi(\vec{k}, \vec{k}_0, \xi)]^2+(4vDk_\parallel)^2
\xi^2 / (3\kappa_\bot k_\bot^2)^2}.
\label{Accurate perpendicular diffusion coefficient-2}
\end{aligned}
\end{equation}
The real part is
\begin{equation}
\begin{aligned}
&\kappa_\bot =\frac{a^2 v^2}{3B_0 ^2 (0)}
\int d^3 k\int d^3 k_0 P_{xx}(\vec{k})h(\vec{k}_0)\\
&\frac{(4/3)\kappa_\bot k_\bot ^2+(k_\parallel v)^2 / (3\kappa_\bot k_\bot^2)
-(2D)^2\xi ^2 / (3\kappa_\bot k_\bot^2)
+\Xi(\vec{k}, \vec{k}_0, \xi)}{[(4/3)\kappa_\bot k_\bot ^2
+(k_\parallel v)^2 / (3\kappa_\bot k_\bot^2) -(2D)^2\xi ^2 / (3\kappa_\bot k_\bot
^2)+\Xi(\vec{k}, \vec{k}_0, \xi)]^2
+(4vDk_\parallel)^2 \xi^2 / (3\kappa_\bot k_\bot^2)^2}.
\label{Accurate perpendicular diffusion coefficient-2}
\end{aligned}
\end{equation}

So far, Equations
(\ref{accurate iteration function})-(\ref{Accurate perpendicular diffusion coefficient-2})
are accurate except for some
common assumptions.

From Equations (\ref{S+ with Lambda+}), (\ref{S- with Lambda-}),
(\ref{accurate iteration function}) and
(\ref{Accurate perpendicular diffusion coefficient-2})
 we can find that
$\Xi(\vec{k}, \vec{k}_0, \xi)$ in Equation
(\ref{Accurate perpendicular diffusion coefficient-2})
is a complicated function of
$\Lambda(\vec{k},\vec{k}_0,\mu)$ which
is the function of $R(\mu)$.
However, $R(\mu)$ is a iteration function by itself.
Therefore, in order to obtain specific expression of $\kappa_\bot$
we have to truncate $R(\mu)$.

\subsection{The conditions of truncating $\boldsymbol{R(\mu)}$}
From the above discussion we can see that $R(\mu)$ has to be truncated.
Comparing to $R_1 (\mu)$ we find
that $R_2 (\mu)$ and $R_3 (\mu)$ can be ignored under some conditions
which can be satisfied by some special cases.
In what follows, we explore these conditions.
And in this paper we only explore the special cases which satisfy these conditions.

Firstly, we compare the relative importance between $R_1(\mu)$ and $R_2(\mu)$.
Let's set $Z_g^*$ and $T_g^*$ as the along-field spatial characteristic scale
and temporal characteristic scale of the anisotropic distribution function $g$
corresponding to the significant magnitude variation.
Here for the purpose of simplification we set the
order of significant magnitude variation as $O(1)$, i.e., $\Delta g=g'\sim O(1)$.
And similarly, $T_F^*$ is set as the temporal characteristic scales
of isotropic distribution function $F$
corresponding to the significant magnitude variation.
And we also approximate the
order of significant magnitude variation as $O(1)$, i.e.,
$\Delta F=F'\sim O(1)$.
Thus, $R_1(\mu)$ and $R_2(\mu)$ can be written as
\begin{eqnarray}
&& R_1(\mu)=\frac{e^{-M(\mu)}}{D_{\mu \mu}}\Bigg[\frac{\mu}{T_F^*}
	\frac{\partial{F'}}
{\partial{t'}}+\frac{1}{T_g^*}
\int_{-1}^{\mu}d\nu \frac{\partial{g'}}{\partial{t'}}
\Bigg], \label{dimensionless R1} \\
&& R_2(\mu)=\frac{e^{-M(\mu)}}{D_{\mu \mu}}
\Bigg[\frac{v}{Z_g^*}\left(\int_{-1}^{\mu}d\nu \nu
\frac{\partial{g'}}{\partial{z'}}-\dfrac{1}
{2}\int_{-1}^{1}d\mu \mu \frac{\partial{g'}}{\partial{z'}}\right)
-\frac{v}{L}\left(\int_{-1}^{\mu}d\nu \nu g'-\dfrac{1}
{2}\int_{-1}^{1}d\mu \mu g'\right) \Bigg],  \label{dimensionless R2}
\end{eqnarray}
where the dimensionless time $t'$ and dimensionless along-field distance $z'$
are used. Then according to the above discussion derivatives terms
in Equation (\ref{dimensionless R1}) can be approximate as
$\partial F'/\partial t'\sim \partial g'/\partial t'\sim O(1)$
\citep[e.g.,][]{Far1994}.
Since pitch-angle cosine $\mu$ belongs to the interval $[-1, 1]$,
so we can approximate the integral terms
in Equation (\ref{dimensionless R2}) as
$[\int_{-1}^{\mu}d\nu \nu
({\partial{g'}}/{\partial{z'}})-(1/2)\int_{-1}^{1}d\mu \mu
({\partial{g'}}/{\partial{z'}})]\sim O(1)$ and
$[\int_{-1}^{\mu}d\nu \nu g'-(1/2)\int_{-1}^{1}d\mu \mu g']\sim O(1)$.
Let's set $T^*$ equal to the maximum value of $T_F^*$ and $T_g^*$.
By comparing Equation (\ref{dimensionless R1})
with Equation (\ref{dimensionless R2})
we can obtain the relation $vT^*\ll Z_g^*$ for the case $L\gg Z^*_g$,
and the relation $vT^*\ll L$ for the case $Z^*_g\gg L$.
If the case $L\sim Z^*_g$ occurs, the relations $vT^*\ll Z_g^*$ and $vT^*\ll L$
have to be satisfied simultaneously.
Therefore, if the condition $vT^*\ll Z_g^*$ and $vT^*\ll L$
are satisfied at the same time, we can find that
$R_2(\mu)$ can be neglected comparing to $R_1(\mu)$ for any case.

In general, if particle motion is completely passive in turbulence, stronger
turbulence
leads to faster change of the distribution function, i.e., stronger turbulence leads
to smaller characteristic temporal scale $T^*$.
It also can be assumed that smaller gradient of anisotropic distribution function
leads to larger spatial characteristic length $Z_g^*$.
Therefore, for relatively strong turbulence, relatively small along-field gradient of
anisotropic distribution function, and not too high energy of particles, the
condition $vT^*\ll Z_g^*$ might be satisfied. The condition $vT^*\ll L$ denotes
the adiabatic focusing characteristic length $L$ is far greater than
the characteristic distance $vT^*$.
Next, we only explore the cases which satisfy the requirements
$vT^*\ll Z_g^*$ and $vT^*\ll L$.
In addition, we assume that perpendicular effects are not much larger than parallel
effects, so we can ignore both $R_2(\mu)$ and $R_3(\mu)$ comparing with $R_1(\mu)$.
Then Equation (\ref{R=R1+R2+R3}) can be simplified as
\begin{equation}
R(\mu)\approx R_1 (\mu)=\frac{e^{-M(\mu)}}{D_{\mu \mu}}\left(\mu \frac{\partial{F}}
{\partial{t}}+\int_{-1}^{\mu}d\nu \frac{\partial{g}}{\partial{t}}\right).
\label{R is equal approximately to R1}
\end{equation}

Combining Equations (\ref{g}) and (\ref{R is equal approximately to R1}) gives
\begin{equation}
\begin{aligned}
g(\mu)=&\left[1-\frac{2e^{M(\mu)}}{\int_{-1}^{1}d\mu e^{M(\mu)}} \right] L
\left(\frac{\partial{F}}{\partial{z}}-\frac{2F}{L}\right)
+e^{M(\mu)}\Bigg[\int_{-1}^{\mu}d\nu \frac{e^{-M(\nu)}}{D_{\nu \nu}}\left(\nu
\frac{\partial{F}}{\partial{t}}
+\int_{-1}^{\nu}d\rho \frac{\partial{g}}{\partial{t}}\right)\\
&-\frac{1}{\int_{-1}^{1}d\mu e^{M(\mu)}}\int_{-1}^{1}d\mu e^{M(\mu)}\int_{-1}^{\mu}
d\nu
\frac{e^{-M(\nu)}}{D_{\nu \nu}}\left(\nu \frac{\partial{F}}
{\partial{t}}+\int_{-1}^{\nu}d\rho \frac{\partial{g}}{\partial{t}}\right)\Bigg].
\label{First-order g}
\end{aligned}
\end{equation}

Next, we operate Fourier transformation on Equation (\ref{First-order g})
and then insert
the result into Equation (\ref{Phi}) to yield
\begin{equation}
\int_{-1}^{1}d\mu_{0} \mu_{0} \int_{0}^{\infty}d t \int d^3 x e^{i \vec{k}\cdot
\vec{x}}
\int_{-1}^{\mu}d\nu R(\nu)
=\frac{1}{D}\int_{-1}^{\mu}d\nu e^{-M(\nu)}\int d^3 x_{0}
e^{i\vec{k}_0\cdot\vec{x}_0}h(\vec{x}_0)+\Phi(\vec{k},\mu)h(\vec{k}_0)
    \label{First-order R}
\end{equation}
with
\begin{eqnarray}
\Phi(\vec{k},\mu)&=&\frac{v}{2D}i k_z \int_{-1}^{\mu}d\nu \frac{e^{-M(\nu)}}
{D_{\nu \nu}}\Bigg\{2\int_{-1}^{\nu}d\rho\rho e^{M(\rho)}\left[\int_{-1}^{\rho}d\sigma
	e^{-
M(\sigma)}-\frac{\int_{-1}^{1}d\mu e^{M(\mu)}\int_{-1}^{\nu}d\nu e^{-M(\nu)}}
{\int_{-1}^{1}d\mu e^{M(\mu)}}\right]\nonumber\\
&{}&-\int_{-1}^{1}d\mu\mu e^{M(\mu)} \left[\int_{-1}^{\mu}d\nu e^{-M(\nu)}-
\frac{\int_{-1}^{1}d\mu e^{M(\mu)}\int_{-1}^{\mu}d\nu e^{-M(\nu)}}{\int_{-1}^{1}
	d\mu
e^{M(\mu)}} \right]\Bigg\}\nonumber\\
&{}&+\frac{k_\bot^2}{2D}\int_{-1}^{\mu}d\nu \frac{e^{-M(\nu)}}{D_{\nu \nu}}
\Bigg\{2\int_{-1}^{\nu}d\rho D_\bot e^{M(\rho)}\left[\int_{-1}^{\rho}d\sigma
	e^{-M(\sigma)}-
\frac{\int_{-1}^{1}d\mu e^{M(\mu)}\int_{-1}^{\mu}d\nu e^{-M(\nu)}}{\int_{-1}^{1}
	d\mu e^{M(\mu)}}\right]\nonumber\\
&{}&-\int_{-1}^{1}d\mu D_\bot e^{M(\mu)} \left[\int_{-1}^{\mu}d\nu e^{-M(\nu)}-
\frac{\int_{-1}^{1}d\mu e^{M(\mu)}\int_{-1}^{\mu}d\nu e^{-M(\nu)}}{\int_{-1}^{1}
	d\mu e^{M(\mu)}} \right]\Bigg \},
\end{eqnarray}
where we assume that $M(\mu)$ and $D_{\mu\mu}(\mu)$ are all independent of the
initial pitch-angle cosine $\mu_0$.
Substituting Equation (\ref{First-order R}) into Equation (\ref{S by R}) yields
\begin{equation}
S(\vec{k},\vec{k}_0,\mu)=\frac{e^{M(\mu)}}{2D}\left[\int_{-1}^{\mu}d\nu
	e^{-M(\nu)}-\frac{\int_{-1}^{1}d\mu e^{M(\mu)}\int_{-1}^{\mu}d\nu
	e^{-M(\nu)}}{\int_{-1}^{1}d\mu
e^{M(\mu)}}\right]h(\vec{k}_0)+\Lambda(\vec{k},\mu)h(\vec{k}_0),
\label{First-order S with k0}
\end{equation}
with
\begin{equation}
\Lambda(\vec{k},\mu)=\frac{1}{2}e^{M(\mu)}\left[\Phi(\vec{k},\mu)-
\frac{\int_{-1}^{1}d\mu
e^{M(\mu)}\Phi(\vec{k},\mu)}{\int_{-1}^{1}d\mu e^{M(\mu)}}\right].
    \label{First-order Lambda}
\end{equation}

Equation (\ref{First-order S with k0}) is the result of first-order iteration of
$R(\mu)$.
For more iterations of $R(\mu)$ we can get more complicated
$S(\vec{k},\vec{k}_0,\mu)$
corresponding to more complicated $\Lambda(\vec{k},\mu)$.
Therefore, if Equation (\ref{R is equal approximately to R1}) is used directly in
Equation (\ref{S by R}), the formula $S(\vec{k},\vec{k}_0,\mu)=S^{(0)}
(\vec{k},\vec{k}_0,\mu)$ can be obtained corresponding to $\Lambda(\vec{k},\mu)=
\Lambda ^{(0)}(\vec{k},\mu)=0$.
Note that we call $S^{(0)}(\vec{k},\vec{k}_0,\mu)$ the zeroth-order
$S(\vec{k},\vec{k}_0,\mu)$.
After inserting Equation (\ref{g}) into Equation (\ref{R=R1+R2+R3}) and using
Equation
(\ref{R is equal approximately to R1}), the first-order $S(\vec{k},\vec{k}_0,\mu)$,
or
$S^{(1)}(\vec{k},\vec{k}_0,\mu)$ can be found corresponding to the first-order
$\Lambda
^{(1)}(\vec{k},\mu)=0$ which is just Equation (\ref{First-order Lambda}).
If we perform n iterations of $R(\mu)$ and use Equation
(\ref{R is equal approximately
to R1}), the nth-order $S(\vec{k},\vec{k}_0,\mu)$, i.e., $S^{(n)}
(\vec{k},\vec{k}_0,\mu)$, can be written as
\begin{equation}
S^{(n)}(\vec{k},\vec{k}_0,\mu)=\Theta (\mu)h(\vec{k}_0)+\Lambda ^{(n-1)}
(\vec{k},\mu)h(\vec{k}_0),
\label{nth-order S}
\end{equation}
with
\begin{equation}
\Theta (\mu)=\frac{e^{M(\mu)}}{2D}\left[\int_{-1}^{\mu}d\nu
e^{-M(\nu)}-\frac{\int_{-1}^{1}d\mu e^{M(\mu)}\int_{-1}^{\mu}d\nu e^{-M(\nu)}}
{\int_{-1}^{1}d\mu e^{M(\mu)}}\right].
\label{nth-order S}
\end{equation}
Here $\Lambda ^{(n-1)}(\vec{k},\mu)$ is a more complicated
formula.
Note that Equation (\ref{S with Lambda(k0)}) denotes the result for infinite
iteration of $R(\mu)$.

\subsection{Truncating $\boldsymbol{\Xi(\vec{k}, \vec{k}_0, \xi)}$}
According to the truncation condition in above subsection, if the approximations
\begin{eqnarray}
[\partial{\Lambda_- (\vec{k}, \vec{k}_0, \mu)}/\partial{\mu}](\mu=0)&\approx&
h(\vec{k}_0)[\partial{\Lambda_-^{(n)} (\vec{k}, \vec{k}_0, \mu)}/\partial{\mu}]
	(\mu=0),\\
\Lambda_+ (\vec{k}, \vec{k}_0, \mu=0)&\approx& h(\vec{k}_0)\Lambda_+^{(n)} (\vec{k},
\mu=0),\\
\Lambda_- (\vec{k}, \vec{k}_0, \mu)&\approx& h(\vec{k}_0)\Lambda_- ^{(n)}(\vec{k}, \mu)
\end{eqnarray}
are used, Equations (\ref{Accurate perpendicular diffusion coefficient}) and
(\ref{accurate iteration function}) becomes
\begin{equation}
\begin{aligned}
\kappa_\bot & =\frac{a^2 v^2}{3B_0 ^2 (0)}\int d^3 k P_{xx}(\vec{k})\\
&\frac{(4/3)\kappa_\bot k_\bot ^2
+(k_\parallel v)^2 / (3\kappa_\bot k_\bot^2) -(2D)^2\xi ^2 / (3\kappa_\bot k_\bot^2)
+\Xi(\vec{k}, \xi)}{[(4/3)\kappa_\bot k_\bot ^2
+(k_\parallel v)^2 / (3\kappa_\bot k_\bot^2) -(2D)^2\xi ^2 / (3\kappa_\bot k_\bot
^2)+\Xi(\vec{k}, \xi)]^2+(4vDk_\parallel)^2 \xi^2 / (3\kappa_\bot k_\bot^2)^2}.
\label{perpendicular diffusion coefficient with nth-order iteration}
\end{aligned}
\end{equation}
with
\begin{equation}
\Xi(\vec{k}, \xi)\approx \frac{2D}{3}\frac{1+2D(\partial{\Lambda_- ^{(n)}(\vec{k},
\mu)}/\partial{\mu})(\mu=0)-2D\xi \Lambda_+ ^{(n)}(\vec{k}, \mu=0)}{
	\left(\xi\cosh \xi -\sinh \xi\right)/\left(\xi^2 \sinh \xi\right)+
	2D\int_{0}^{1}d\mu\mu\Lambda_- ^{(n)}\left(\vec{k},\mu\right)},
\label{nth-order iteration}
\end{equation}
where the normalization condition $\int d^3 k_0 h(\vec{k}_0)=1$ is used. Equation
(\ref{perpendicular diffusion coefficient with nth-order iteration}) with formula
(\ref{nth-order iteration}) is the nth-order result of $\kappa_\bot (\xi)$.

The expression $\Xi(\vec{k}, \xi)$ is a very complicated function. For mathematical
tractability in this paper we only consider the zeroth-order $\kappa_\bot (\xi)$
corresponding to $\Lambda ^{(0)}(\vec{k}, \mu=0)=0$ and $\partial{\Lambda_- ^{(0)}
(\vec{k}, \mu)}/\partial{\mu})(\mu=0)=0$ in Equation (\ref{nth-order iteration}).
Then Equation (\ref{perpendicular diffusion coefficient with nth-order iteration})
can be simplified as
\begin{equation}
\begin{aligned}
\kappa_\bot &=\frac{a^2 v^2}{3B_0 ^2 (0)}\int d^3 k P_{xx}(\vec{k})\\
&\frac{(4/3)\kappa_\bot k_\bot ^2
+(k_\parallel v)^2 / (3\kappa_\bot k_\bot^2) -(2D)^2\xi ^2 / (3\kappa_\bot k_\bot^2)
+\Xi(\xi)}{[(4/3)\kappa_\bot k_\bot ^2+(k_\parallel v)^2 / (3\kappa_\bot k_\bot^2)
-(2D)^2\xi ^2 / (3\kappa_\bot k_\bot
^2)+\Xi(\xi)]^2+(4vDk_\parallel)^2 \xi^2 / (3\kappa_\bot k_\bot^2)^2}.
\label{perpendicular diffusion coefficient with hyperbolic function}
\end{aligned}
\end{equation}
with
\begin{equation}
\Xi(\xi)=\frac{2D}{3}\frac{\xi^2 \sinh \xi}{\xi \cosh \xi - \sinh\xi}
\end{equation}
From the latter equation we can see that $\kappa_\bot (\xi)$ is a complicated
function of $\xi$. And for weak limit $|\xi|\ll 1$ Equation
(\ref{perpendicular diffusion coefficient with hyperbolic function}) can be written
as
\begin{equation}
\begin{aligned}
\kappa_\bot &\approx\frac{a^2 v^2}{3B_0 ^2 (0)}
\int d^3 k P_{xx}(\vec{k})\\
&\frac{1}{\left[2D+\frac{4}{3}\kappa_\bot k_\bot ^2
+\frac{(k_\parallel v)^2}{3\kappa_\bot k_\bot^2}\right]
+\left[\frac{2D}{15} -\frac{(2D)^2}{3\kappa_\bot k_\bot^2}
+\frac{\frac{(4vDk_\parallel)^2}{(3\kappa_\bot k_\bot^2)^2}}
{\left[2D+\frac{4}{3}\kappa_\bot k_\bot ^2+\frac{(k_\parallel v)^2}
{3\kappa_\bot k_\bot^2}\right] +\frac{2D}{15}\xi ^2-\frac{(2D)^2\xi ^2}
{3\kappa_\bot k_\bot^2}}\right]\xi^2}\\
&\approx \frac{a^2 v^2}{3B_0 ^2 (0)}\int d^3 k P_{xx}(\vec{k})\\
&\frac{1}{\left[2D+\frac{4}{3}\kappa_\bot k_\bot ^2
+\frac{(k_\parallel v)^2}
{3\kappa_\bot k_\bot^2}\right]+\left[\frac{2D}{15}
-\frac{(2D)^2}{3\kappa_\bot k_\bot^2}+\frac{\frac{(4vDk_\parallel)^2}
{(3\kappa_\bot k_\bot^2)^2}} {2D+\frac{4}{3}\kappa_\bot k_\bot ^2
+\frac{(k_\parallel v)^2}{3\kappa_\bot k_\bot^2}}\right]\xi^2}.
\label{perpendicular diffusion coefficient for weak limit}
\end{aligned}
\end{equation}
The latter equation can be rewritten as series expansion
\begin{equation}
\kappa_\bot=\kappa_\bot^{[0]}\left[1-\frac{\kappa_\bot^{[1]}}
{\kappa_\bot^{[0]}}\xi^2+\frac{\kappa_\bot^{[2]}}{\kappa_\bot^{[0]}}\xi^4
-\frac{\kappa_\bot^{[3]}}{\kappa_\bot^{[0]}}\xi^6+\cdots\cdots +(-1)^n \frac{\kappa_\bot^{[n]}}
{\kappa_\bot^{[0]}}\xi^{2n}+\cdots\cdots \right]
\end{equation}
with
\begin{eqnarray}
&&\kappa_\bot^{[0]}=\frac{a^2 v^2}{3B_0 ^2 (0)}\int d^3 k
\frac{P_{xx}(\vec{k})}{2D+\frac{4}{3}\kappa_\bot k_\bot^2
+\frac{(k_\parallel v)^2}{3\kappa_\bot k_\bot^2}}\\
&& \kappa_\bot^{[1]}=\frac{a^2 v^2}{3B_0 ^2 (0)}\int d^3 k
\frac{P_{xx}(\vec{k})}{2D+\frac{4}{3}\kappa_\bot k_\bot^2
+\frac{(k_\parallel v)^2}{3\kappa_\bot k_\bot^2}}
\frac{\frac{2D}{15}-\frac{(2D)^2}{3\kappa_\bot k_\bot^2}
+\frac{(\frac{4Dk_\parallel v}{3\kappa_\bot k_\bot^2})^2}
{2D+\frac{4}{3}\kappa_\bot k_\bot^2
+\frac{(k_\parallel v)^2}{3\kappa_\bot k_\bot^2}}}{2D+\frac{4}{3}
\kappa_\bot k_\bot^2
+\frac{(k_\parallel v)^2}{3\kappa_\bot k_\bot^2}}\\
&& \kappa_\bot^{[2]}=\frac{a^2 v^2}{3B_0 ^2 (0)}\int d^3 k
\frac{P_{xx}(\vec{k})}{2D+\frac{4}{3}\kappa_\bot k_\bot^2
+\frac{(k_\parallel v)^2}{3\kappa_\bot k_\bot^2}}
\left[\frac{\frac{2D}{15}-\frac{(2D)^2}{3\kappa_\bot k_\bot^2}
+\frac{(\frac{4Dk_\parallel v}{3\kappa_\bot k_\bot^2})^2}
{2D+\frac{4}{3}\kappa_\bot k_\bot^2
+\frac{(k_\parallel v)^2}{3\kappa_\bot k_\bot^2}}}{2D+\frac{4}{3}
\kappa_\bot k_\bot^2+\frac{(k_\parallel v)^2}{3\kappa_\bot k_\bot^2} }\right]^2 \\
&&\kappa_\bot^{[3]}=\frac{a^2 v^2}{3B_0 ^2 (0)}\int d^3 k
\frac{P_{xx}(\vec{k})}{2D+\frac{4}{3}\kappa_\bot k_\bot^2
+\frac{(k_\parallel v)^2}{3\kappa_\bot k_\bot^2}}
\left[\frac{\frac{2D}{15}-\frac{(2D)^2}{3\kappa_\bot k_\bot^2}
+\frac{(\frac{4Dk_\parallel v}{3\kappa_\bot k_\bot^2})^2}
{2D+\frac{4}{3}\kappa_\bot k_\bot^2
+\frac{(k_\parallel v)^2}{3\kappa_\bot k_\bot^2}}}{2D+\frac{4}{3}
\kappa_\bot k_\bot^2+\frac{(k_\parallel v)^2}{3\kappa_\bot k_\bot^2} }\right]^3
\end{eqnarray}
\hspace{2cm}$\cdots\cdots$
\begin{eqnarray}
 &&\kappa_\bot^{[n]}=\frac{a^2 v^2}{3B_0 ^2 (0)}\int d^3 k
\frac{P_{xx}(\vec{k})}{2D+\frac{4}{3}\kappa_\bot k_\bot^2
+\frac{(k_\parallel v)^2}{3\kappa_\bot k_\bot^2}}
\left[\frac{\frac{2D}{15}-\frac{(2D)^2}{3\kappa_\bot k_\bot^2}
+\frac{(\frac{4Dk_\parallel v}{3\kappa_\bot k_\bot^2})^2}
{2D+\frac{4}{3}\kappa_\bot k_\bot^2
+\frac{(k_\parallel v)^2}{3\kappa_\bot k_\bot^2}}}{2D+\frac{4}{3}
\kappa_\bot k_\bot^2+\frac{(k_\parallel v)^2}{3\kappa_\bot k_\bot^2} }\right]^n
\end{eqnarray}
\hspace{2cm}$\cdots\cdots$

Obviously, when parameter $\xi$ tends to zero Equation
(\ref{perpendicular diffusion coefficient for weak limit})
tends to the result of \citet{Shalchi2010}.
The perpendicular and parallel diffusion coefficient
$\kappa_\bot^{(0)}$ and $\kappa_\parallel^{(0)}$
for uniform background magnetic field
have been explored deeply \citep[e.g.,][]{Shalchi2009}.
So it is convenient to explore the influence
of adiabatic focusing on perpendicular diffusion
if the modifying factors
to perpendicular diffusion coefficient $\kappa_\bot (\xi)$,
which is the function of $\kappa_\bot^{(0)}$ and $\kappa_\parallel^{(0)}$,
can be obtained.

{\bf \subsection{Power function expanding formula of $\boldsymbol{\kappa_\bot (\xi)}$}}

Perpendicular diffusion coefficient $\kappa_\bot(\xi)$
can be
expanded as power function of $\xi$ as the following
\begin{equation}
\kappa_\bot (\xi)=\kappa_\bot ^{(0)}(1+a_1\xi + a_2\xi^2
	+ \cdots\cdots),
	\label{perpendicular diffusion coefficient with adiabatic focusing}
\end{equation}
where $\kappa_\bot ^{(0)}$ is the perpendicular diffusion coefficient for uniform
background field, and $a_1$, $a_2$, $\cdots$ are
the modifying factors to perpendicular diffusion coefficient
introduced by adiabatic focusing.
If $\xi \rightarrow 0$ the background magnetic field tends to be uniform.
For $\xi =0$ only $\kappa_\bot ^{(0)}$ is retained in
formula (\ref{perpendicular diffusion coefficient with adiabatic focusing}).
On the other hand, the following parallel diffusion coefficient with along-field
adiabatic focusing was already derived in previous papers
\citep[see, e.g.,][]{ShalchiEA2013, Litvinenko2012a,  HeEA2014}
\begin{equation}
\kappa_\parallel (\xi)=\kappa_\parallel ^{(0)}\left(1+b_2 \xi^2 +
\cdots\cdots\right),
\label{parallel diffusion coefficient with adiabatic focusing}
\end{equation}
where $\kappa_\parallel ^{(0)}$ is the parallel diffusion coefficient for uniform
background field, and $b_2=-1/15$.

Inserting Equations
(\ref{perpendicular diffusion coefficient with adiabatic focusing})
and (\ref{parallel diffusion coefficient with adiabatic focusing}) into Equation
(\ref{perpendicular diffusion coefficient for weak limit}) yields
\begin{eqnarray}
a_1&=&0, \\
a_2&=&\frac{b_2\frac{a^2 v^4}{9B_0 ^2 (0)\kappa_\parallel ^{(0)}}
\int d^3 k \frac{P_{xx}(\vec{k})}{\Omega^2}+\frac{a^2v^2}{3B_0^2(0)}
\int d^3 k \frac{P_{xx}(\vec{k})}{\Omega^2}[\frac{v^4}{27k_\bot^2\kappa_\bot^{(0)}
(\kappa_\parallel ^{(0)})^2}-\frac{v^2}{45\kappa_\parallel ^{(0)}}
-\frac{k_\parallel ^2 v^6}{9\Omega
k_\bot ^4 (\kappa_\bot^{(0)})^2(\kappa_\parallel^{(0)})^2}
]}{\kappa_\bot^{(0)}-\frac{a^2v^2}{3B_0^2(0)}\int d^3 k \frac{P_{xx}(\vec{k})}
{\Omega^2}\left[\frac{(k_\parallel v)^2}{3k_\bot^2\kappa_\bot^{(0)}}
-\frac{4}{3}k_\bot^2\kappa_\bot^{(0)} \right]},
\label{a2}
\end{eqnarray}
with
\begin{eqnarray}
\kappa_\bot ^{(0)}=\frac{a^2 v^2}{3B_0 ^2 (0)}
\int d^3 k  \frac{P_{xx}(\vec{k})}{\Omega}, \\
\end{eqnarray}
and
\begin{eqnarray}
\Omega&=&\frac{v^2}{3\kappa_\parallel ^{(0)}}+\frac{(k_\parallel v)^2}{3k_\bot ^2
\kappa_\bot ^{(0)}} +\frac{4}{3}k_\bot ^2 \kappa_\bot ^{(0)},
\end{eqnarray}

It is noteworthy that Equation
(\ref{parallel diffusion coefficient with adiabatic focusing}) is only
applicable to the cases
in which perpendicular diffusion is not considered and $D_{\mu\mu}$ is
independent of adiabatic focusing.
So it is more appropriate to derive the formulas of
$D_{\mu\mu}$ and
$\kappa_\parallel$
with adiabatic focusing and perpendicular diffusion effects,
and then to re-deduce Equation
(\ref{a2}). we would
reserve this as our tasks in the future.

\section{MODIFYING FACTOR ${\bf a_2}$}
For mathematical tractability we only
explore the modifying factor $a_2$
for the two-component model (slab+2D, Matthaeus et al. 1990).
For slab component contribution it is more convenient to start from Equation
(\ref{perpendicular diffusion coefficient for weak limit}). We can easily
find that the slab transport is subdiffusive, i.e., $\kappa_\bot^{slab}=0$.
So we only consider the 2D-component contribution.
For pure two-dimensional model Equation (\ref{a2}) can be simplified as
\begin{eqnarray}
a_2= \frac{\frac{a^2v^4}{27B_0^2(0)\kappa_\parallel ^{(0)}}
}{\kappa_\bot^{(0)}+\frac{4a^2v^2}{9B_0^2(0)}\int d^3 k
\frac{P_{xx}^{2D}(\vec{k})}{\Omega^2}k_\bot^2\kappa_\bot^{(0)}}W,
\label{a2w}
\end{eqnarray}
with
\begin{eqnarray}
W&=&\int d^3 k \frac{P_{xx}^{2D}(\vec{k})}{\Omega^2}\left[\frac{v^2}
{3k_\bot^2\kappa_\bot^{(0)}\kappa_\parallel ^{(0)}}-\frac{2}{5}\right], \\
\label{w1}
\Omega&=&\frac{v^2}{3\kappa_\parallel ^{(0)}}
+\frac{4}{3}k_\bot ^2 \kappa_\bot ^{(0)}.
\end{eqnarray}

In this paper the
following tensor of the two-dimensional (2D) magnetic turbulence is used
\begin{equation}
P_{lm}^{2D}(\vec{k})=g^{2D}(k_\bot)\frac{\delta(k_\parallel)}{k_\bot}
	\left[\delta_{l m}-\frac{k_l k_m}{k^2}\right], \hspace{1cm} l,m=x,y,
	\label{2D-tensor}
\end{equation}
here we use the spectra of the two-dimensional modes $g^{2D}(k_\bot)$
\begin{equation}
g^{2D}(k_\bot)=\frac{D(s,q)}{2\pi}l_{2D}\delta B_{2D}^2
\begin{cases}
0,                     &   \text{$k_\bot < L_{2D}^{-1}$}\\
(k_\bot l_{2D})^q,     &   \text{$L_{2D}^{-1}<k_\bot<l_{2D}^{-1}$}\\
(k_\bot l_{2D})^{-s},  &   \text{$l_{2D}^{-1}<k_\bot< k_{cut}$},    \label{2D-mode}
\end{cases}
\end{equation}
where $q$ is the energy range index, $s$ is the inertial range index, $L_{2D}$ is box size,
$l_{2D}$ is the turnover scale, $k_{cut}$ is the inertial range cut-off wave number
and $D(s,q)$ is the normalized factor. The spectrum is correctly normalized
for $q>-1$ and $s>1$.
The more detailed explanation can be found in \citet{Shalchi2011c}.

From Equation (\ref{a2w}) we can find
that the sign of $a_2$ is determined by formula $W$.
In what follows,
we explore the sign of formula $W$.
From formulas
(\ref{2D-tensor}) and (\ref{2D-mode}) we
can obtain
\begin{equation}
W=\frac{81l_{2D}^3}{32v^2\left(\lambda_\bot^{(0)}\right)^2}D(s,q)\delta B_{2D}^2
\left[\int_\eta ^1 dx x^{q-2} \frac{(5\varepsilon/2-x^2)}{(3\varepsilon/4+x^2)^2}
+\int_1 ^{k_{cut}}dx x^{-s-2}
\frac{(5\varepsilon/2-x^2)}{(3\varepsilon/4+x^2)^2}\right],
\label{w2}
\end{equation}
here formulas $\varepsilon
=v^2l_{2D}^2/(3\kappa_\parallel^{(0)}\kappa_\bot^{(0)})
=3l_{2D}^2/(\lambda_\parallel^{(0)}\lambda_\bot^{(0)})$,
$x=k_\bot l_{2D}$,
$\eta=l_{2D}/L_{2D}$ and $\Omega
=v^2/(3\kappa_\parallel ^{(0)})+4k_\bot^2\kappa_\bot^{(0)}/3$ are used.

\subsection{The case $\boldsymbol{\varepsilon=3l_{2D}^2/(\lambda_\parallel^{(0)}\lambda_\bot^{(0)})\rightarrow 0}$}
This condition $\varepsilon\rightarrow 0$, i.e., $l_{2D}^2
\ll \lambda_\parallel^{(0)}\lambda_\bot^{(0)}$,
which denotes the product of parallel
and perpendicular mean free paths $\lambda_\parallel ^{(0)}$
and $\lambda_\bot ^{(0)}$ is far
greater than the turnover scale $l_{2D}$.
And high energy charged particles satisfy this condition.
For this case $W$ becomes
\begin{equation}
W\approx -\frac{81l_{2D}^4}{32v^2(\lambda_\bot^{(0)})^2}D(s,q)\delta B_{2D}^2
\left[\frac{1-\eta^{q-3}}{q-3}+\frac{1-k_{max}^{-s-3}}{s+3}\right].
\label{w3}
\end{equation}
since $k_{max}^{-s-3}\ll 1$, the latter equation can be simplified as
\begin{equation}
W\approx -\frac{81l_{2D}^4}{32v^2(\lambda_\bot^{(0)})^2}D(s,q)\delta B_{2D}^2
\left[\frac{1-\eta^{q-3}}{q-3}+\frac{1}{s+3}\right].
\label{w3}
\end{equation}
For $q>3$ we can find $\eta^{q-3}\ll 1$, then we can get
\begin{equation}
W\approx -\frac{81l_{2D}^4}{32v^2(\lambda_\bot^{(0)})^2}D(s,q)\delta B_{2D}^2
\frac{s+q}{(q-3)(s+3)}<0.
\label{w4}
\end{equation}
For $-1<q<3$ Equation (\ref{w3}) can be simplified as
\begin{equation}
W\approx -\frac{81l_{2D}^4}{32v^2(\lambda_\bot^{(0)})^2}D(s,q)\delta B_{2D}^2
\frac{\eta^{q-3}(s+3)+(3-q)}{(3-q)(s+3)}<0.
\label{w5}
\end{equation}
In both special case (\ref{w4}) and (\ref{w5}) we find that $W<0$, then $a_2<0$.
Thus, we can see that for high energy particles the modified factor $a_2$
make perpendicular diffusion coefficient decrease.

\subsection{The case $\boldsymbol{\varepsilon=3l_{2D}^2/(\lambda_\parallel^{(0)}\lambda_\bot^{(0)})\rightarrow \infty}$}
In this case the condition
$\lambda_\parallel^{(0)}\lambda_\bot^{(0)}\ll l_{2D}^2$
corresponding to low energy particles should be satisfied.
And we can derive
the following equation from equation (\ref{w2})
\begin{equation}
W\approx \frac{15l_{2D}^2\lambda_\parallel^{(0)}}{4v^2\lambda_\bot^{(0)}}D(s,q)
\delta B_{2D}^2 \left[\frac{1-\eta^{q-1}}{q-1}+\frac{1-k_{max}^{-s-1}}{s+1}\right].
\label{w6}
\end{equation}
since $k_{max}^{-s-1}\ll 1$, we can get from the latter equation
\begin{equation}
W\approx \frac{15l_{2D}^2\lambda_\parallel^{(0)}}{4v^2\lambda_\bot^{(0)}}D(s,q)
\delta B_{2D}^2 \left[\frac{1-\eta^{q-1}}{q-1}+\frac{1}{s+1}\right].
\label{w6}
\end{equation}
For $q>1$ we get $\eta^{q-1}\ll 1$
and the latter formula can be simplified as
\begin{equation}
W\approx \frac{15l_{2D}^2\lambda_\parallel^{(0)}}{4v^2\lambda_\bot^{(0)}}D(s,q)
\delta B_{2D}^2 \frac{s+q}{(q-1)(s+1)}>0.
\label{w7}
\end{equation}
For $-1<q<1$ we can obtain from formula (\ref{w6})
\begin{equation}
W\approx \frac{15l_{2D}^2\lambda_\parallel^{(0)}}{4v^2\lambda_\bot^{(0)}}D(s,q)
\delta B_{2D}^2 \left[\frac{\eta^{q-1}}{1-q}+\frac{1}{s+1}\right]>0.
\label{w8}
\end{equation}
Clearly we find that for low energy particles the modified factor $a_2$
make perpendicular diffusion coefficient increase.

In summary, the specific modifying factor $a_2$
to perpendicular diffusion coefficient
appears different forms for different particle energy.
However, it is independent of energy range index $q$ and
the sign of adiabatic focusing length.
Considering the discussion in Section 4.1,
it is might only the results obtained in Subsection 5.2 that
can satisfy the application scope of this paper.

\section{DISCUSSION AND CONCLUSIONS}
In this paper by employing the Unified NonLinear Transport (UNLT) theory in SH2010
and
the method of HS2014 we explore the perpendicular diffusion of energetic charged
particles in spatially varying background magnetic field, and the perpendicular
diffusion coefficient with along-field adiabatic focusing is obtained.
From this formula we find that along-field adiabatic focusing
can introduce significant
modification to the particle transport across the mean magnetic field.

Previous articles demonstrated
that the parallel diffusion is decreased by along-field
adiabatic focusing, and parallel diffusion coefficient is independent of
the sign of adiabatic focusing
\citep{BeeckEA1986,BieberEA1990,Kota2000,Shalchi2011a,
Litvinenko2012a,Litvinenko2012b,ShalchiEA2013,HeEA2014}.
In this paper we find that for isotropic scattering and weak adiabatic limit
the perpendicular diffusion coefficient for non-uniform background field
is independent of
the sign of adiabatic focusing length,
which is similar to parallel diffusion coefficient.
The perpendicular diffusion coefficient can be indicated as
power series of adiabatic focusing characteristic quantity $\xi$
and all of the modifying factors are shown as the functions
of perpendicular and parallel diffusion coefficients
for uniform background magnetic field.
For two-component model we find that the first order modifying factor
is equal to zero and the sign of the second
one is determined by the energy of particles.
For low energy particles the second order modifying factor is positive,
and for high energy particles the modifying factor
is negative.

But it must been emphasized that the perpendicular diffusion coefficient derived in
this
paper with three requirements: The first one is that the turbulence is relatively
strong, the along-field gradient of
anisotropic distribution function is small, and the energy of particles is not too
high, so that $vT^*\ll Z_g^*$ can be satisfied.
The second one is that the perpendicular diffusion
effect is not much higher than the parallel one.
The third one is adiabatic focusing length is very large, i.e.,
$vT^*\ll L$.
So the formulas derived in this paper might be only applicable to the special case
of Section 5.2 corresponding to low energy particles.
It is also noted that the specific formulas shown in this paper are only for the
lowest order iteration results.
By combining Equations (\ref{T with HS2014}) and
(\ref{Originally perpendicular diffusion coefficient with T})
perpendicular diffusion coefficient $\kappa_\bot (\xi)$
with along-field adiabatic focusing can also be obtained.
This is our future task.
In fact, because we only explore
the weak adiabatic focusing limit $L\rightarrow \infty$ in this paper,
formula
$B_0 (z)=B_0 (0) e^{-z/L}\approx B_0 (0)$ is established approximately.
So the results derived by employing the approximation
$B_0 (z)\approx B_0 (0)$ is also approximately valid.

\acknowledgments
We are partly supported by grants NNSFC 41125016,  NNSFC 41574172, and NNSFC
41374177.


\begin{thebibliography}{}

\bibitem[Beeck \& Wibberenz(1986)]{BeeckEA1986}Beeck, J., \& Wibberenz, G. 1986, \apj,
	311, 437
\bibitem[Bieber \& Burger(1990)]{BieberEA1990}Bieber, J. W., \& Burger, R. A. 1990,
	\apj, 348, 597
\bibitem[Corrsin(1959)]{Corrsin1959}Corrsin, S. 1959,
in Progress Report on Some Turbulent Diffusion Research, Atmospheric Diffusion and Air Pollution, Adv. Geophys., Vol. 6,
ed. F. Frenkiel \& P. Sheppard (New York: Academic), 161
\bibitem[Earl(1976)]{Earl1976}Earl, J. A. 1976, \apj, 205, 900
\bibitem[Fay(1994)]{Far1994}Fay, J. A. 1994, Fluid Mechanics (Cambridge: MIT Press)
\bibitem[Ferrand et al.(2014)]{FerrandEA14}Ferrand, G., Danos, R. J.,
Shalchi, A., Safi-Harb, S., Edmon, P., \& Mendygral, P. 2014, \apj, 792, 133
\bibitem[He \& Schlickeiser(2014)]{HeEA2014}He, H.-Q., \& Schlickeiser, R. 2014,
	\apj, 792, 85
\bibitem[Hussein \& Shalchi(2014)]{HusseinEA2014}Hussein, M., \& Shalchi, A. 2014,
	\mnras, 444, 2676
\bibitem[Green(1951)]{Green1951}Green, M. S. 1951, JChPh, 19, 1036
\bibitem[K\'ota(2000)]{Kota2000}K\'ota, J. 2000, \jgr, 105, 2403
\bibitem[Kubo(1957)]{Kubo1957}Kubo, R. 1957, JPSJ, 12, 570
\bibitem[Litvinenko(2012a)]{Litvinenko2012a}Litvinenko, Y. E. 2012a, \apj, 752, 16
\bibitem[Litvinenko(2012b)]{Litvinenko2012b}Litvinenko, Y. E. 2012b, \apj, 745, 62
\bibitem[Matthaeus et al.(1990)]{MatthaeusEA90}Matthaeus, W. H., Goldstein, M. L., \& Roberts, D. A. 1990, \jgr, 95, 20673
\bibitem[Matthaeus et al.(2003)]{MatthaeusEA2003}Matthaeus, W. H., Qin, G., Bieber,
	J.
W., \& Zank, G. P. 2003, \apj, 590, L53
\bibitem[Ng \& Reames(1994)]{NgAReames94}Ng, C. K., \& Reames, D. V. 1994, \apj,
424, 1032
\bibitem[Parker(1965)]{Parker65}Parker, E. N. 1965, \planss, 13, 9
\bibitem[Parker(1963)]{Parker63}Parker, E. N. 1963, Interplanetary Dynamical
Processes (New York: Interscience)
\bibitem[{Qin} {et~al.}(2005)]{QinEA05}
{Qin}, G., {Zhang}, M., {Dwyer}, J. R., {Rassoul}, H. K., \& {Mason}, G. M. 2005, \apj, 627, 562
\bibitem[Qin \& Shalchi(2014)]{QinAShalchi14}Qin, G., \& Shalchi, A. 2014,
Applied Physics Research, 6, 1
\bibitem[Qin \& Zhang(2014)]{QinEA2014}Qin, G., \& Zhang, L.-H. 2014, \apj, 787, 12
\bibitem[Roelof(1969)]{Roelof1969}Roelof, E. C. 1969, in Lectures in High Energy
Astrophysics, ed. H. \"Ogelmann \& J. R. Wayland(NASA SP-199: Washington, DC:
		NASA), 111
\bibitem[Schlickeiser(2002)]{Schlickeiser2002}Schlickeiser, R. 2002, Cosmic Ray
Astrophysics (Berlin: Springer)
\bibitem[Schlickeiser \& Shalchi (2008)]{SchlickeiserEA2008}Schlickeiser, R.,
	\& Shalchi, A. 2008, \apj, 686, 292
\bibitem[Schlickeiser \& Jenko (2010)]{SchlickeiserEA2010}Schlickeiser, R., \& Jenko,
	F. 2010, J. Plasma Physics, 76, 317
\bibitem[Shalchi et al. (2004)]{ShalchiEA2004}Shalchi, A., Bieber, J. W., Matthaeus,
	W. H., \& Qin, G. 2004, \apj, 616, 617
\bibitem[Shalchi(2009)]{Shalchi2009}Shalchi, A. 2009, Nonlinear Cosmic Ray
Diffusion Theories, Astrophysics and Space Science Library, Vol. 362 (Berlin: Springer)
\bibitem[Shalchi et al.(2009)]{ShalchiEA2009}Shalchi, A., Skoda, T., Tautz, R. C., \& Schlickeiser, R. 2009, \aap, 507, 589
\bibitem[Shalchi(2010)]{Shalchi2010}Shalchi, A. 2010, ApJL, 720, L127
\bibitem[Shalchi et al. (2010)]{ShalchiEA2010}Shalchi, A., Li, G., \& Zank, G. P. 2010, Ap\&SS, 325, 99
\bibitem[Shalchi(2011a)]{Shalchi2011a}Shalchi, A. 2011a, \apj, 728, 113
\bibitem[Shalchi(2011b)]{Shalchi2011b}Shalchi, A. 2011b, Plasma Phys. Control. Fusion, 53, 074010
\bibitem[Shalchi(2011c)]{Shalchi2011c}Shalchi, A. 2011c, Contrib. Plasma Phys., 51, 920
\bibitem[Shalchi(2013)]{Shalchi2013}Shalchi, A. 2013, \apss, 344, 187
\bibitem[Shalchi \& Danos(2013)]{ShalchiEA2013}Shalchi, A., \& Danos, R. J. 2013, \apj, 765, 153
\bibitem[Shalchi \& Hussein(2014)]{ShalchiEA2014}Shalchi, A., \& Hussein, M. 2014, \apj, 794, 56
\bibitem[Shalchi(2015)]{Shalchi2015}Shalchi, A. 2015, AdSpR, 56, 1264
\bibitem[Shalchi(2016)]{Shalchi2016}Shalchi, A. 2016, AdSpR, 57, 431
\bibitem[Shalchi(2017)]{Shalchi2017}Shalchi, A. 2017, Physics of plasmas, 24, 050702
\bibitem[Taylor(1922)]{Taylor1922}Taylor, G. I. 1922, Proc. London Math. Soc., 20, 196
\bibitem[Tautz \& Shalchi(2011)]{TautzEA2011}Tautz, R. C., \& Shalchi, A. 2011, \apj, 735, 92
\bibitem[Wang \& Qin (2016)]{WangAQin16}Wang, Y., \& Qin, G. 2016, \apj,  820, 61
\bibitem[Zank et al.(2000)]{ZankEA00}Zank, G. P., Rice, W. K. M., \& Wu, C. C. 2000,
\jgr, 105, 25079
\bibitem[Zhao et al.(2014)]{ZhaoEA14}Zhao, L.-L., Qin, G., Zhang, M., \& Heber, B.
2014, \jgr, 119, 1493

\end{thebibliography}
\end{document}